\documentclass[bib]{statapress}

\usepackage[crop,newcenter,frame]{pagedims}

\usepackage{amsmath}

\usepackage{graphicx}

\usepackage{sj}

\usepackage{epstopdf}

\usepackage{epsfig}

\usepackage{stata}

\usepackage{shadow}

\sjsetissue{$\#$}{$\#$}{$\#$}{$2020$}

\usepackage{bm}

\begin{document}


\inserttype[]{article}

\author{Giovanni Cerulli}{
Giovanni Cerulli\\IRCrES-CNR\\Rome, Italy\\giovanni.cerulli@ircres.cnr.it
}

\title[Machine Learning in Stata]{Machine Learning using Stata/Python}

\maketitle

\begin{abstract}
We present two related Stata modules, \texttt{r\_ml\_stata} and \texttt{c\_ml\_stata}, for fitting 
popular Machine Learning (ML) methods both in a regression and a classification setting. Using the recent Stata/Python integration platform (sfi) of Stata 16, these commands provide hyper-parameters' optimal tuning via K-fold cross-validation using greed search.  More specifically, they make use of the Python Scikit-learn API to carry out both cross-validation and outcome/label prediction.

\end{abstract}

\section[Introduction]{Introduction}

Machine learning (ML) has emerged as a leading data science approach in many fields of human activities, including business, engineering, medicine, advertisement, and scientific research. Placing itself in the intersection between statistics, computer science, and artificial intelligence, ML's main objective is turning information into valuable knowledge by ``letting the data speak'', limiting the model's prior assumptions, and promoting a model-free philosophy. Relying on algorithms and computational techniques, more than on analytic solutions, ML targets Big Data and complexity reduction, although sometimes at the expense of results' interpretability (Hastie, Tibshirani, and Friedman, 2001; Varian, 2014). 

Unlike other software such as R, Python, Matlab, and SAS, Stata has not dedicated built-in packages for fitting ML algorithms, if one excludes the Lasso package of Stata 16 (StataCorp, 2019). Recently, however, the Stata community has developed some popular ML routines that Stata users can suitably exploit. Among them, we mention Schonlau (2005) implementing a Boosting Stata plugin; Guenther and Schonlau (2016) providing a command fitting Support Vector Machines (SVM); Ahrens, Hansen, and Schaffer (2020) setting out the \texttt{lassopack}, a set of commands for model selection and prediction with regularized regression; and Schonlau and Zou (2020) recently providing a module for the random forest algorithm.  All these are valuable packages to carry out  popular ML algorithms within Stata, but they lack uniformity and comparability as they show up as stand--alone routines poorly integrated one another.  

To pursuit generality, uniformity and comparability one should rely on a software platform able to suitably integrate most of the mainstream ML methods. Python, for example, has powerful platforms to carry out both ML and deep--learning algorithms (Raschka and Mirjalili, 2019). Among them, the most popular are Scikit--learn for fitting a large number of ML methods, and Tensorflow and Keras for fitting neural network and deep--learning techniques more in general. These make of Python, that is also a freeware software,  probably the most effective and complete software for ML and deep--learning available within the community. 

The Stata 16 release has introduced a useful Stata/Python interface API, the Stata Function Interface (SFI) module, that allows users to interact Python's capabilities with core features of Stata. The module can be used interactively or in do--files and ado--files.

Using the Stata/Python integration interface, this paper presents two related Stata modules, \texttt{r\_ml\_stata} and \texttt{c\_ml\_stata} wrapping Python functions for fitting popular Machine Learning (ML) methods both in a regression and a classification setting. Both commands provide hyper-parameters' optimal tuning via $K$--fold cross--validation by implementing greed search.  More specifically, they make use of the Python Scikit-learn API to carry out both cross-validation and outcome/label prediction.

Why the Stata community would need these commands? I see three related answers to this question. First, Stata users who are willing to apply ML methods would benefit from these commands as they can allow them to perform ML without time--consuming investment in learning other software. Second, these commands provide users with standard Stata returns and variables' generation that can be further used in subsequent Stata coding within the same do--file. This makes the user's workflow smoother, less error-prone, and faster. Third, these commands represent a sort of a case--study showing the potential of integrating Stata and Python. To my experience, indeed, Stata is unbeatable compared to Python when it comes to performing intensive data management and manipulation, descriptive statistics, and immediate graphing, while Python is more suitable for its large set of implemented ML algorithms, for an all--encompassing availability of libraries, for computing speed and sophisticated graphing (not only 3D, but also deep contour plots, and images' reproduction).   

The structure of the paper is as follows. Section \ref{sec:introML} presents a brief introduction to machine learning, where subsection \ref{sec:basicsML} sets out the basics, subsection \ref{sec:tuning} the optimal tuning procedure via cross-validation, and subsection \ref{sec:learn_arch} the learning methods and architecture. 
Section \ref{sec:syntax} and \ref{sec:syntax2} present the syntax of \texttt{r\_ml\_stata} and \texttt{c\_ml\_stata}. For users to familiarize with these commands, section \ref{sec:app1} illustrates a simple step--by--step application for fitting a regression tree, while sections \ref{sec:app2} and \ref{sec:app3} show how to apply the previous commands to a real dataset for either classification and regression purposes. Section \ref{sec:concl} concludes the paper.

\section{A brief introduction to machine learning} \label{sec:introML}

\subsection{The basics of machine learning} \label{sec:basicsML}

Machine learning is the branch of artificial intelligence mainly focused on statistical prediction (Boden, 2018). The literature distinguishes between supervised and unsupervised learning, referring to a setting where the outcome variable is known (supervised) or unknown (unsupervised). In statistical terms, supervised learning coincides with a regression or classification setup, where regression represents the case in which the outcome variable is numerical, and classification the case in which it is categorical. In contrast, unsupervised learning deals with unlabeled data, and its main concern is that of generating a categorical variable via proper clustering algorithms. Unsupervised learning can thus be encompassed within statistical cluster analysis.

In this article, we focus on supervised machine learning\footnote{This section and the subsequent one are based on the book of Hastie, Tibshirani, and Friedman (2001). }. We are thus interested in predicting either numerical variables (regression) or label classes (classification) as a function of $p$ predictors (or features) that may be both quantitative or qualitative. From a statistical point of view, we want to fit the conditional expectation of $Y$, the outcome, on a set of $p$ predictors, $X$, starting from the following population equation:

\begin{equation} \label{Eq: eq1}
Y = f(X_{1}, \dots, X_{p}) + \epsilon
\end{equation}
where $\epsilon$ is an error term with mean equal to zero and finite variance. By taking the expectation over $X$, and assuming that  $E(\epsilon|X)=0 $, we have:      
\begin{equation} \label{Eq: eq2}
Y = E(Y|X) + \epsilon = f(X_{1}, \dots, X_{p}) + \epsilon
\end{equation} 
 that is:
\begin{equation} \label{Eq: eq3}
E(Y|X) = f(X_{1}, \dots, X_{p}) 
\end{equation} 
The conditional expectation $f(\cdot)$ represents a function mapping predictors and expected outcomes. The main purpose of supervised machine learning is that of fitting Eq. (\ref{Eq: eq3}) with the aim of reducing as much as possible the prediction error coming up when one wants to predict actual outcome variables. 
The prediction error can be generally defined as $e=Y-\hat{f}(X)$, with $\hat{f}(\cdot)$ indicating a specific ML method. Generally, ML scholars focus on minimizing the mean square error (MSE), defined as $MSE=E[Y-\hat{f}(X)]^2$. However, while the in--sample MSE (the so--called ``training--MSE'') is generally affected by overfitting (thus going to zero as the model's degrees of freedom - or complexity - raises up), the out--of--sample MSE (also known as ``test-MSE'') has the property to be a convex function of model complexity, and thus characterized by an optimal level of complexity.    

\begin{figure}[ht]
\begin{center}
\includegraphics[width=10cm]{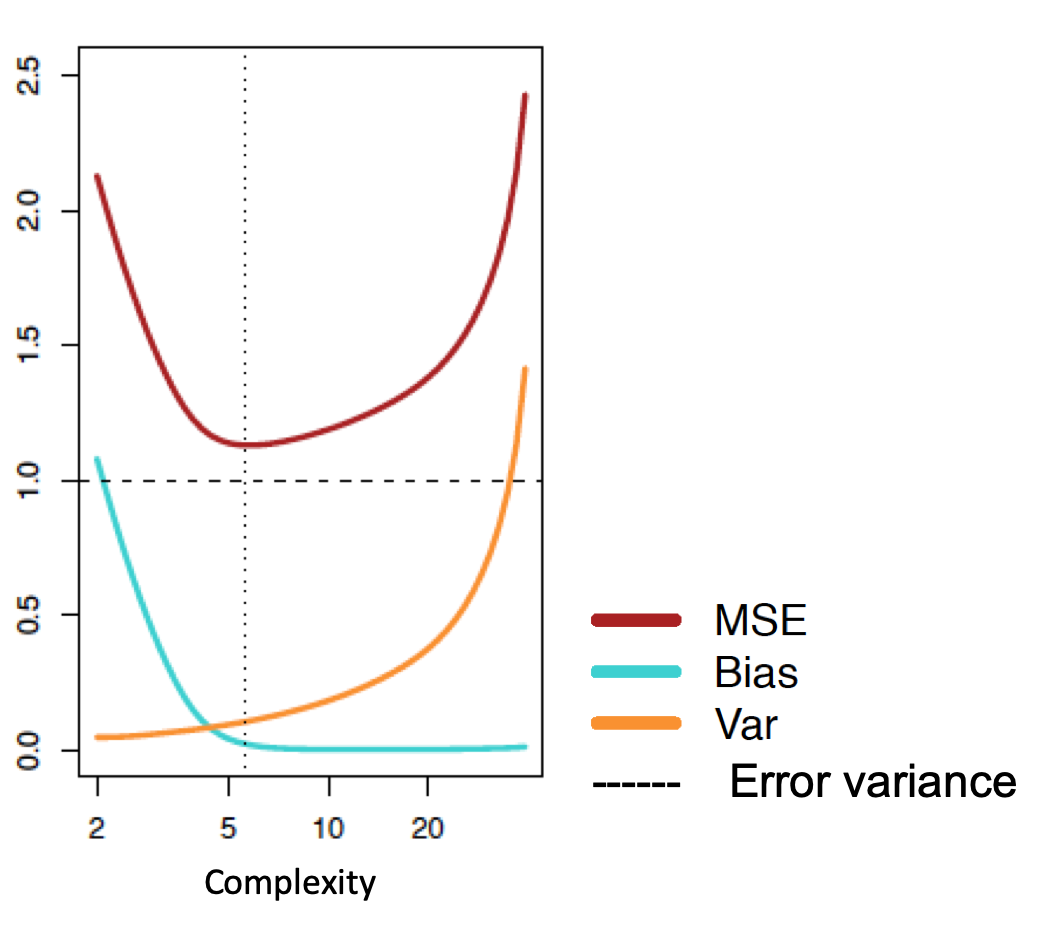}
\end{center}
\caption{Trade-off between bias and variance as functions of model complexity.}
\label{fig:fig1}
\end{figure}

It can be proved that the test-MSE, when evaluated at one out--of--sample observation (also called ``new instance'' in the ML jargon), can be decomposed into three components -- variance, bias squared, and irreducible error -- as follows:  

   \begin{equation} \label{Eq: eq2}
E[(Y-\hat{f}(X_{0})]^2 = Var(\hat{f}(X_{0}) + [Bias(\hat{f}(X_{0})]^2+Var(\epsilon)
\end{equation} 
where $X_{0}$ is a new instance, that is, an observation that did not participate in producing the ML fit $\hat{f}(\cdot)$. Figure \ref{fig:fig1} shows a graphical representation of the pattern of the previous quantities as functions of model complexity. It is immediate to see that, as long as model complexity increases, the bias decreases while the variance increases monotonically. Because of this, the test--MSE sets out a parabola-shaped pattern which allows us for minimizing it at a specific level of model complexity. This is the optimal model tuning whenever complexity is measured by a specific hyper-parameter $\lambda$. In the figure, the irreducible error variance represents a constant lower bound of the test--MSE. It is not possible to overcome this minimum test--MSE, as it depends on the nature of the data generating process (intrinsic unpredictability of the phenomenon under analysis).   

In the classification setting, the MSE is meaningless as in this case we have class labels and not numerical values. For classification purposes, the correct objective function to minimize is the (test) mean classification error (MCE) defined as:

\begin{equation} \label{Eq: eq4}
MCE = E[I(y\neq \hat{C}(X))]
\end{equation} 
where $I(\cdot)$ is an index function, and $\hat{C}(X)$ the fitted classifier. As in the case of the MSE, it can be proved that the training--MCE overfits the data when model complexity increases, while the test--MCE allows us to find the optimal model's complexity. Therefore, the graph of figure \ref{fig:fig1} can be likewise extended also to the case of the test--MCE. 

\subsection{Optimal tuning via cross-validation} \label{sec:tuning}
Finding the optimal model complexity, parameterized by a generic hyper--parameter $\lambda$, is a computational task. There are basically three ways to tune an ML model:

\begin{itemize}
\item Information criteria
\item $K$--fold cross--validation
\item Bootstrap
\end{itemize}

Information criteria are based on goodness--of--fit formulas that adjust the training error by penalizing too complex models (i.e., models characterized by large degrees of freedom). Traditional information criteria  comprise the Akaike criterion (AIC) and the Bayesian information criterion (BIC), and can be applied to both linear and non--linear models (probit, logit, poisson, etc.). Unfortunately, the information criteria are valid only for linear or generalized linear models, i.e. for parametric regression. They cannot be computed for nonparametric methods like -- for example -- tree--based or nearest neighbor regressions. 

For nonparametric models, the test--error can be estimated via computational techniques, more specifically, by resampling methods. Boostrap -- resampling with replacement from the original sample -- could in theory be a practical solution, provided that the original dataset is used as validation dataset and the bootstrapped ones as training datasets. Unfortunately, the bootstrap has the limitation of generating observation overlaps between the test and the training datasets, as about two-thirds of the original observations appear in each bootstrap sample. This occurrence undermines its use to validate out--of--sample an ML procedure. 

Cross--validation is the workhorse of the test--error estimation in machine learning. The idea is to randomly divide the initial dataset into $K$ equal--sized portions called \textit{folds}. This procedure suggests to leave out fold $k$ and fit the model to the other ($K$-1) folds (wholly combined) to then obtain predictions for the left-out $k$--th fold. This is done in turn for each fold $k = 1, 2,\dots, K$, and then the results are combined by averaging the ($K$-1) estimates of the error. The cross--validation procedure can be carried out as follows: 

\begin{itemize}

\item Split randomly the initial dataset into $K$ folds denoted as  $G_{1},G_{2},\dots,G_{K}$, where $G_{k}$ refers to part $k$. Assume that there are $n_{k}$ observations in fold $k$. If $N$ is a multiple of $K$, then $n_{k} = n/K$.

\item For each fold $k = 1, 2,\dots, K$, compute:
$$MSE_{k} = \sum_{i \in C_{k}} (y_{i}-\hat{y_{i}})^2/n_{k}$$ 
where  $\hat{y_{i}}$ is the fit for observation $i$, obtained from the dataset with fold $k$ removed. 
 
\item Compute: 
$$ CV_{K} = \sum_{i=1}^{K}\frac{n_{k}}{n}MSE_{k} $$
that is the average of all the out--of--sample $MSE$ obtained fold--by--fold.

\end{itemize}
Observe that, by setting $K = n$, we obtain the $n$--fold or leave-one out cross-validation (LOOCV). Also, as $CV_{K}$ is an estimation the true test--error, estimating its standard error can be useful to provide a confidence interval and thus a measure of test--accuracy's uncertainty.

For ML classification purposes, finally, the cross-validation procedure follows the same line, except for considering the MCE in place of the MSE.

\subsection{Learning methods and architecture} \label{sec:learn_arch}

A learner $L_{j}$ is a mapping from the set $[X, \theta, \lambda_{j},f_{j}(\cdot)]$ to an outcome $y$, where $X$ is the matrix of features, $\theta$ a vector of estimation parameters, $\lambda_{j}$ a vector of tuning parameters, and $f_{j}(\cdot)$ an algorithm taking as inputs $X$, $\theta$, and $\lambda_{j}$.  Differently from the members of the Generalized Linear Models (GLM) family (linear, probit or multinomial regressions are classical examples), that are highly parametric and are not characterized by tuning parameters, machine learning models -- such as local--kernel, nearest--neighbor,  or tree--based regressions -- may be highly nonparametric and characterized by one or more hyper-parameters $\lambda_{j}$ which may be optimally chosen to minimize the \emph{test prediction error}, i.e. the out--of--sample predicting accuracy of the learner, as stressed in the previous section.

A detailed description of all the available machine learning methods is beyond the scope of this paper.
Table \ref{tab:tab1}, however, sets out the most popular machine learning algorithms proposed in the literature along with the associated tuning hyper--parameters.  

\begin{table}
\centering
\small
\begin{tabular}{llll}
\hline \\
\textbf{ML method}              & \textbf{Parameter 1 }     & \textbf{Parameter 2 }  & \textbf{Parameter 3}  \\
\hline \\
\textit{Linear Models and GLS}  & N. of covariates             &                      &                     \\
\textit{Lasso}                            & Penalization coefficient         &                      &                     \\
\textit{Elastic-Net }                  & Penalization coefficient         & Elastic parameter    &                     \\
\textit{Nearest-Neighbor }     & N. of neighbors              &                      &                     \\
\textit{Neural Network}         & N. of hidden layers          & N. of neurons    &                     \\
\textit{Trees}                             & N. of leaves     &                      &                     \\
\textit{Boosting}                       & Learning parameter               & N. of bootstraps & N. of leaves          \\
\textit{Random Forest }           & N. of features for splitting & N. of bootstraps & N. of leaves          \\
\textit{Bagging }                       & Tree-depth                       & N. of bootstraps &                     \\
\textit{Support Vector Machine} & C                                & $\Gamma$                &                     \\
\textit{Kernel regression}        & Bandwidth                        & Kernel function          &                     \\
\textit{Piecewise regression}   & N. of knots                  &                      &                     \\
\textit{Series regression}          & N. of series terms           &                      &               \\
\hline     
\end{tabular}
\caption{Main machine learning methods and associated tuning hyper--parameters.}
\label{tab:tab1}
\end{table}

\begin{figure}[ht]
\centering
\includegraphics[width=14cm]{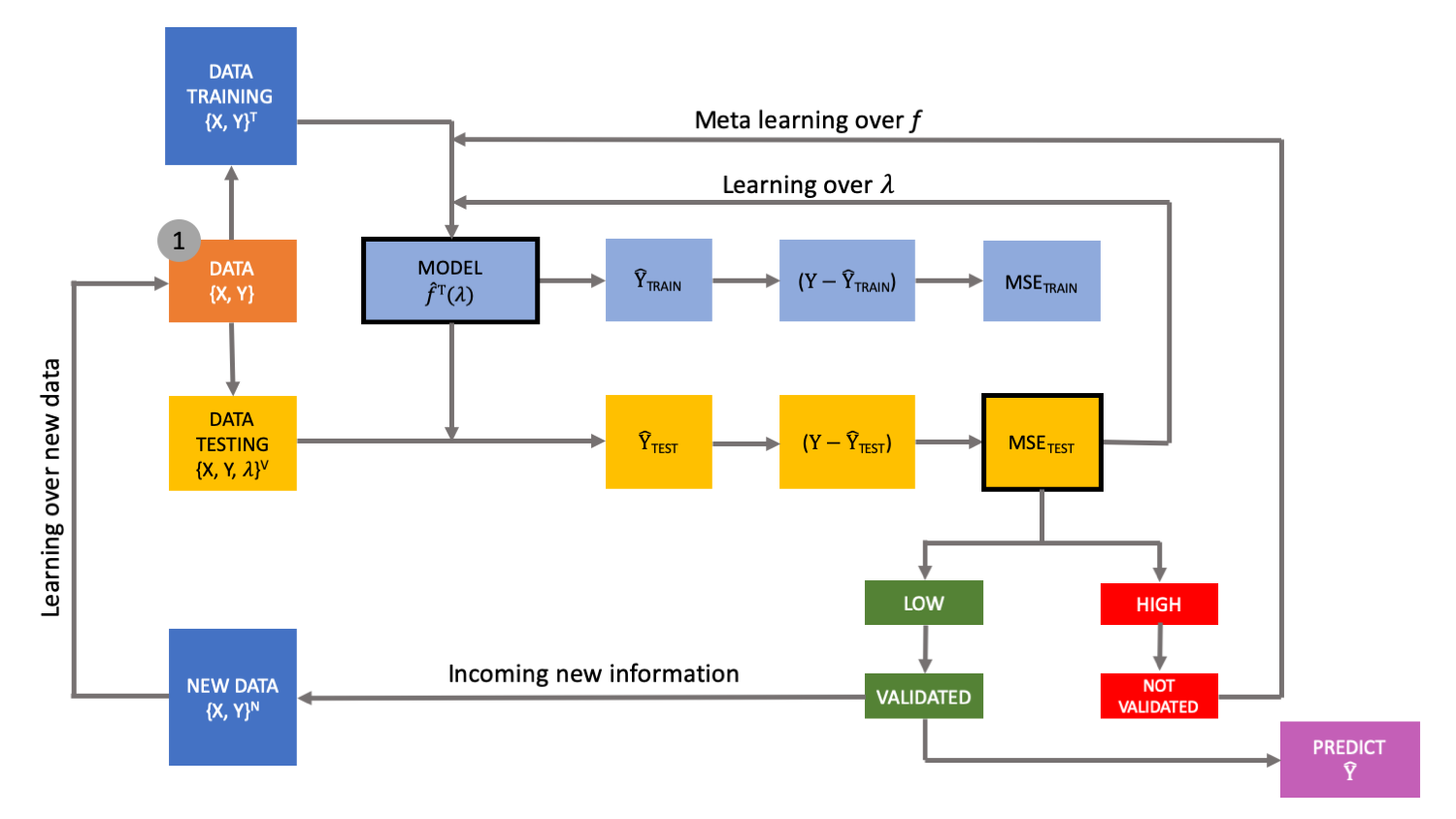}
\caption{The meta--learning machine architecture.}
\label{fig:fig2}
\end{figure}

A combined use of these methods can produce a computational architecture (i.e., a virtual learning machine) enabling to increase statistical prediction and its estimated precision (Van der Laan, Polley and Hubbard, 2007). Figure \ref{fig:fig2} presents the learning architecture proposed by Cerulli (2020). This framework is made of three linked learning processes: (i) the learning over the tuning parameter $\lambda$, (ii) the learning over the algorithm $f(\cdot)$, and (iii) the learning over new additional information. The departure is in point 1, from where we set off assuming the availability of a dataset $[X,y]$.   

The first learning process aims at selecting the optimal tuning parameter(s) for a given algorithm $f_{j}(\cdot)$. As seen above, ML scholars typically do it using $K$-fold cross--validation to draw test--accuracy (or, equivalently, test--error) measures and related standard deviations.    

At the optimal $\lambda_{j}$, one can recover the largest possible prediction accuracy for the learner $f_{j}(\cdot)$. Further prediction improvements can be achieved only by learning from other learners, namely, by exploring other $f_{j}(\cdot)$, with $j=1, \dots, M$ (where $M$  is the number of learners at hand). 

Figure \ref{fig:fig1} shows the training estimation procedure that corresponds to the light blue sequence of boxes leading to the $MSE_{TRAIN}$ which is, \textit{de facto}, a dead-end node, being the training error plagued by \textit{overfitting}. 

Conversely, the yellow sequence leads  to the $MSE_{TEST}$, which is informative to take correct decisions about the predicting quality of the current learner. At this node, the analyst can compare the current $MSE_{TEST}$ with a benchmark one (possibly, pre--fixed), and conclude whether to predict using the current learner, or explore alternative learners in the hope of increasing predictive performance. If the level of the current prediction error is too high, the architecture would suggest to explore other learners.     

In the ML literature, learning over learners is called \emph{meta learning}, and entails an exploration of the out--of--sample performance of alternative algorithms $f_{j}(\cdot)$  with the goal of identifying one behaving better than the those already explored (Van der Laan and Rose, 2011). For each new $f_{j}(\cdot)$, our architecture finds an optimal tuning parameter and a new estimated accuracy (along with its standard deviation). The analyst can either explore the entire bundle of alternatives and finally pick--up the best one, or decide to select the first learner whose accuracy is larger than the benchmark. Either cases are automatically run by this virtual machine. 

The third final learning process concerns the availability of new information, via additional data collection. This induces a reiteration of the initial process whose final outcome can lead to choose a different algorithm and tuning parameter(s), depending on the nature of the incoming information. 

As final step, one may combine predictions of single optimal learners into one single super--prediction (\emph{ensemble learning}). What is the advantage of this procedure? As an average, this method cannot provide the largest accuracy possible. However, as sums of i.i.d. random variables have smaller variance than the single addends, the benefit consists of a smaller predictive uncertainty (Zhou, 2012).


\section{Syntax}

\subsection{Syntax for \texttt{r\_ml\_stata}} \label{sec:syntax}

The command \texttt{r\_ml\_stata} fits machine learning regression algorithms. It considers as the main inputs a continuous response variable $y$ (i.e. the \textit{depvar}), a series of predictors (or features) in \textit{varlist} explaining the $y$, plus a series of options.

\begin{stsyntax}
\dunderbar{}r\_ml\_stata
    \depvar\
    \varlist\ ,    
    {\tt mlmodel({\it modeltype\/})}
    {\tt out\_sample({\it filename\/})}
    {\tt in\_prediction({\it name\/})}
    {\tt out\_prediction({\it name\/})}
    {\tt cross\_validation({\it name\/})}
    {\tt seed({\it integer\/})}
    \optional{,
   {\tt save\_graph\_cv({\it name})}
    }
\end{stsyntax}

\hangpara
\textit{depvar} is a numerical variable. Missing values are not allowed.

\hangpara
\textit{varlist} is a list of numerical variables representing the features. When a feature is categorical, please generate the categorical dummies related to this feature. As the command does not do it by default, it is user's responsibility to generate the appropriate dummies.  Missing values are not allowed.

\vspace{0.5cm}

\textbf{Options}

\hangpara
{\tt mlmodel({\it modeltype\/})} specifies the machine learning algorithm to be estimated.
\textit{modeltype} takes the following options: 
\texttt{elasticnet} (Elastic--net), \texttt{tree} (Regression tree), \texttt{randomforest} (Bagging and Random forests), \texttt{boost} (Boosting), \texttt{nearestneighbor} (Nearest Neighbor), \texttt{neuralnet} (Neural network), \texttt{svm} (Support vector machine).

\hangpara
{\tt out\_sample({\it filename\/})} requests to provide a new dataset in filename containing the new instances over which estimating predictions. This dataset contains only features.

\hangpara
{\tt in\_prediction({\it name\/})} requires to specify a name for the file that will contain in--sample predictions.

\hangpara
{\tt out\_prediction({\it name\/})} requires to specify a name for the file that will contain out--sample predictions, those obtained from the option {\tt out\_sample({\it filename\/}).} 

\hangpara
{\tt cross\_validation({\it name\/})} requires to specify a name for the dataset that will contain cross--validation results.  The command uses $K$--fold cross--validation, with $K$=10 by default.

\hangpara
{\tt seed({\it integer\/})} requests to specify a integer seed to assure replication of same results.

\hangpara
{\tt save\_graph\_cv({\it name})} allows to obtain the cross-validation optimal tuning graph drawing the pattern of both train and test accuracy.

\vspace{0.5cm}

\textbf{Returns}

\hangpara
\texttt{r\_ml\_stata} returns into e-return scalars (if numeric) or macros (if string) the ``optimal hyper-parameters'', the ``optimal train accuracy'', the ``optimal test accuracy'', and the ``standard error of the optimal test accuracy'' obtained via cross--validation.

\vspace{0.5cm}

\textbf{Remarks}

\hangpara
Missing values in both the outcome and the list of features are not allowed.  Before running this command, please check whether your dataset presents missing values and delete them. 

\hangpara
To run this program you need to have both Stata 16 and Python (from version 2.7 onwards) installed.  Also, the  Python packages Scikit-learn, Pandas, Numpy, and Scipy must be installed before running the command.


\subsection{Syntax for \texttt{c\_ml\_stata}} \label{sec:syntax2}

The command \texttt{c\_ml\_stata} fits machine learning classification algorithms. It considers as the main inputs a categorical response variable $y$ (i.e. the \textit{depvar}), a series of predictors (or features)  in \textit{varlist} explaining the $y$, plus a series of options.

\begin{stsyntax}
\dunderbar{}c\_ml\_stata
    \depvar\
    \varlist\ ,    
    {\tt mlmodel({\it modeltype\/})}
    {\tt out\_sample({\it filename\/})}
    {\tt in\_prediction({\it name\/})}
    {\tt out\_prediction({\it name\/})}
    {\tt cross\_validation({\it name\/})}
    {\tt seed({\it integer\/})}
    \optional{,
   {\tt save\_graph\_cv({\it name})}
    }
\end{stsyntax}

\hangpara
\textit{depvar} is a numerical discrete dependent variable representing the different classes. It is recommended to re--code this variable so to take values $[1,2,...,M]$ in a $M$-class setting.  For example, if the outcome is binary taking values $[0,1]$, remember to record it so to take values $[1,2]$.  Missing values are not allowed.

\hangpara
\textit{varlist} is a list of numerical variables representing the features. When a feature is categorical, please generate the categorical dummies related to this feature. As the command does not do it by default, it is user's responsibility to generate the appropriate dummies.  Missing values are not allowed.

\vspace{0.5cm}

\textbf{Options}

\hangpara
{\tt mlmodel({\it modeltype\/})} specifies the machine learning algorithm to be estimated.
\textit{modeltype} takes the following options: 
\texttt{tree} (Tree-based classification), \texttt{randomforest} (Bagging and Random forests), \texttt{boost} (Boosting), \texttt{regularizedmultinomial} (Regularized multinomial), \texttt{nearestneighbor} (Nearest neighbor), \texttt{neuralnet} (Neural network), \texttt{naivebayes} (Na\"{i}ve Bayes), \texttt{svm} (Support vector machine).

\hangpara
{\tt out\_sample({\it filename\/})} requests to provide a new dataset in filename containing the new instances over which estimating predictions. This dataset contains only features.

\hangpara
{\tt in\_prediction({\it name\/})} requires to specify a name for the file that will contain in--sample predictions.

\hangpara
{\tt out\_prediction({\it name\/})} requires to specify a name for the file that will contain out--sample predictions, those obtained from the option {\tt out\_sample({\it filename\/}).} 

\hangpara
{\tt cross\_validation({\it name\/})} requires to specify a name for the dataset that will contain cross--validation results.  The command uses $K$--fold cross--validation, with $K$=10 by default.

\hangpara
{\tt seed({\it integer\/})} requests to specify a integer seed to assure replication of same results.

\hangpara
{\tt save\_graph\_cv({\it name})} allows to obtain the cross-validation optimal tuning graph drawing the pattern of both train and test accuracy.

\vspace{0.5cm}

\textbf{Returns}

\hangpara
\texttt{c\_ml\_stata} returns into e-return scalars (if numeric) or macros (if string) the ``optimal hyper-parameters'', the ``optimal train accuracy'', the ``optimal test accuracy'', and the ``standard error of the optimal test accuracy'' obtained via cross-validation.

\hangpara
\texttt{c\_ml\_stata} provides model predictions both as predicted labels and as predicted probabilities.

\vspace{0.5cm}

\textbf{Remarks}

\hangpara
Missing values in both outcome and varlist are not allowed.  Before running this command, please check whether your dataset presents missing values and delete them. 

\hangpara
To run this program you need to have both Stata 16 and Python (from version 2.7 onwards) installed.  Also, the  Python packages Scikit-learn, Pandas, Numpy, and Scipy must be installed before running the command.

\section{Application 1: fitting a regression tree} \label{sec:app1}

This section presents an illustrative application of the use of both \texttt{r\_ml\_stata} within a cross--section data structure. It is thought to allow users to become familiar with the use of this module (the use of \texttt{c\_ml\_stata} follows a similar procedure). To begin with, I show how to implement step--by--step a regression tree.

\begin{itemize}
\item \textbf{Step 1}. Before starting, install Python (from version 2.7 onwards), and the Python packages \texttt{Scikit--learn}, \texttt{Numpy}, \texttt{Pandas}, and \texttt{Scipy}. If you need a help on how to install Python and its packages look at the Python webpage\footnote{Specifically, look at the Python installation page: https://realpython.com/installing-python.}. 

\item \textbf{Step 2}. Once you have Python installed in your machine, you need to install the Stata ML command:

\begin{stverbatim}
\begin{verbatim}
. ssc install r_ml_stata 
\end{verbatim}
\end{stverbatim}

and look at the documentation file of the command to explore its syntax:

\begin{stverbatim}
\begin{verbatim}
. help  r_ml_stata 
\end{verbatim}
\end{stverbatim}

\item \textbf{Step 3}. The command requires to provide a dataset with no missing values. It is user's responsibility to assure this. We can thus load the training dataset prepared for this example:

\begin{stverbatim}
\begin{verbatim}
. use "r_ml_stata_data_example"
\end{verbatim}
\end{stverbatim}

This dataset contains one target variable ($y$) and 13 features $(x_{1}, x_{2}, ... , x_{13})$. All variables are numerical and thus suitable for running our regression tree. 

Before running the command, however, a testing dataset must be provided. This is a dataset made of the same features of the training one, but with \textit{new} instances. Observe that this dataset must neither contain missing values, nor include the target variable ($y$). In this example, we prepared a testing dataset called \texttt{r\_ml\_stata\_data\_new\_example}. This dataset must be in the same directory of the training dataset.

\item \textbf{Step 4}. There are now all the ingredients to run our regression tree. We simply run these lines of code in Stata:

\begin{stverbatim}
\begin{verbatim}
 . r_ml_stata y x1-x13 , mlmodel(tree) in_prediction("in_pred")  ///
   cross_validation("CV")  out_sample("r_ml_stata_data_new_example")  ///   
   out_prediction("out_pred") seed(10) save_graph_cv("graph_cv")
\end{verbatim}
\end{stverbatim}

where the syntax has this meaning:

\begin{itemize}
\item The argument \texttt{tree} tells Stata to run a tree--regression. Other options are available (see the help-file).
\item The argument \texttt{in\_pred} tells Stata to generate a dataset \texttt{in\_pred.dta} containing the in--sample predictions of the estimated model. They are the predictions for only the training dataset.
\item The argument \texttt{out\_pred} tells Stata to generate a dataset \texttt{out\_pred.dta} containing the out--of--sample predictions of the estimated model. They are predictions only for the testing dataset.
\item The argument \texttt{r\_ml\_stata\_data\_new\_example} tells Stata to use this one as testing dataset.
\item The seed is necessary to replicate the same results and must be an integer. 
\item The argument \texttt{graph\_cv} tells Stata to save the cross--validation results graph in your current directory. 

\end{itemize}

\item \textbf{Step 5}. In order to access the main results of the command thus run, you can look at the command's \texttt{ereturn} list by typing:

\begin{stverbatim}
\begin{verbatim}

. ereturn list
---------------------------------------
scalars:

         e(OPT_LEAVES) =  4

         e(TEST_ACCURACY) =  .2027650052251946

         e(TRAIN_ACCURACY) =  .8911061692860425

         e(BEST_INDEX) =  3
         
         e(SE_TEST_ACCURACY) =  .742518604039193
---------------------------------------

\end{verbatim}
\end{stverbatim}

We can observe that the cross--validated (CV) optimal number of leaves (namely, the tree optimal final nodes) is 4, the CV optimal train accuracy is 0.89, while the CV optimal test accuracy is much smaller, i.e. 0.20.  The accuracy measure is the share of the total outcome variance explained by the model (it is closely similar to an adjusted R-squared). Finally, the standard error of the test accuracy is equal to 0.74. 

\item \textbf{Step 6}. Interestingly, the command provides also a graphical representation of the 10--fold cross--validation results carried out. The graph in figure \ref{fig:fig3}, in fact, shows that the optimal grid search index is 3. At this index, corresponding to 4 leaves, the test accuracy is the maximum over the grid. It is also useful to observe the \textit{overfitting} pattern of the train accuracy going to one (maximum accuracy) as long as the model complexity increases. This phenomenon justifies ML focus on just the test accuracy which shows, in this graph, a clear variance--bias trade--off.      

\begin{figure}[ht]
\centering
\includegraphics[width=14cm]{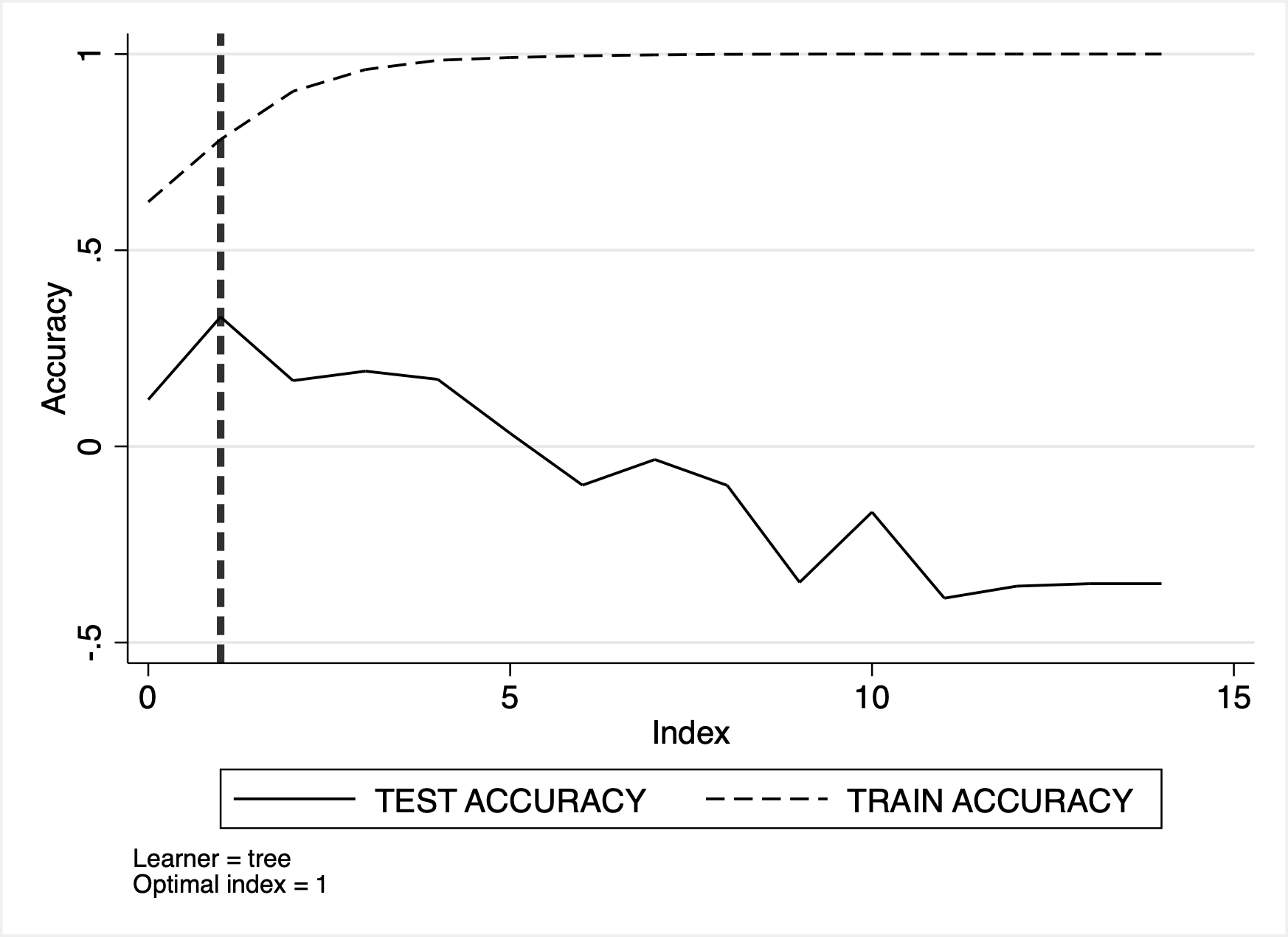}
\caption{Cross-validation graph for a tree regression.}
\label{fig:fig3}
\end{figure}

\item \textbf{Step 7}. We can go even deeper into the understanding of the cross-validation results, by opening the CV results' dataset \texttt{CV.dta} and list its content: 

\begin{stverbatim}
\begin{verbatim}

. use CV , clear

. list

     +--------------------------------------------+
     | index   mean_tr~e   mean_tes~e   std_tes~e |
     |--------------------------------------------|
  1. |     0   .46707705   -.06167094   .39788509 |
  2. |     1   .70630139    .19044095    .4556592 |
  3. |     2   .82658573     .0736554   .81239835 |
  4. |     3   .89110617    .20276501    .7425186 |
  5. |     4    .9226751    .03514619    1.106288 |
     |--------------------------------------------|
  6. |     5   .94706553    .05360043   .80043332 |
  7. |     6    .9643006    .08797992    .8920612 |
  8. |     7   .97651532    .03100165   .89505782 |
  9. |     8   .98468366   -.10475403   .99490224 |
 10. |     9   .99037207    .04775965   .91827191 |
     |--------------------------------------------|
 11. |    10   .99419849   -.10484366   .98651178 |
 12. |    11   .99669445   -.20313959   1.2912981 |
 13. |    12   .99800409   -.10410404   .97689497 |
 14. |    13   .99881627    .07493419   .85811843 |
 15. |    14   .99929196   -.03926511   .91545614 |
     +--------------------------------------------+
     
\end{verbatim}
\end{stverbatim}

Results show, by every grid index, the train accuracy, the test accuracy, and the standard error of the test accuracy estimated over the 10--fold runs. The standard error is important, as it measures the precision we obtain when estimating the test accuracy. In this example, at the optimal index (i.e., 3), the test accuracy's standard error is 0.74, which should be compared with those obtained from other ML algorithms. This means that the choice of the ML model to employ for prediction purposes should ponder not only the level of the achieved test accuracy, but also its standard error.  

\item \textbf{Step 8}. Finally, we can have a look at the out-of-sample predictions. This can be done by opening and listing the \texttt{out\_pred} dataset:

\begin{stverbatim}
\begin{verbatim}

. use out_pred , clear

. list

     +-------------------+
     | index   out_sam~d |
     |-------------------|
  1. |     0   21.629744 |
  2. |     1   16.238961 |
  3. |     2   21.629744 |
  4. |     3   16.238961 |
  5. |     4   16.238961 |
     |-------------------|
  6. |     5   16.238961 |
  7. |     6   16.238961 |
  8. |     7   16.238961 |
  9. |     8   16.238961 |
 10. |     9   21.629744 |
     |-------------------|
 11. |    10   27.427273 |
     +-------------------+

\end{verbatim}
\end{stverbatim}

We observe that the predictions are made of only three values $[21.62, 16.23, 27.42]$ corresponding to three out of the four optimal terminal tree leaves. Graphically it represents a step--function (omitted for the sake of brevity).  

\end{itemize}

\section{Application 2: ML classification} \label{sec:app2}
In this section, we show an illustrative application of \texttt{c\_ml\_stata} using the popular \texttt{auto} dataset. We intend to guess whether a ``new'' car is a ``foreign'' or ``domestic'' one based on a series of characteristics, including price, number of repairs, weight, etc. Our goal is to provide a comparison of the accuracy performed by the classifiers available through \texttt{c\_ml\_stata}. This is in tune with the learning architecture outlined in section \ref{sec:learn_arch}, as it can provide guidance over the choice of the proper learner to use.  In order to carry out this analysis, we proceed in three steps: the first bunch of code cleans and prepares the datasets (training and test) to use for fitting \texttt{c\_ml\_stata} in a correct way; the second bunch of code fits all the learners available to the data; a third and final part of the code yields a forest plot for visualizing and comparing learners' test accuracy and standard deviation.

We start by setting out the code for data cleaning and preparation:
      
\begin{stverbatim}
\begin{verbatim}
********************************************************************************
* DATA CLEANING AND PREPARATION
******************************************************************************** 
* Clean the detaset eliminating all labels and missing values
clear all
cd "/Users/giocer/Dropbox/Stata_Python/Paper_c_r_ml_stata"
sysuse auto , clear
label drop _all 
drop make
qui reg _all
keep if e(sample)

* Standardize the features 
global X "price mpg rep78 headroom trunk weight length turn displacement gear_ratio"
global XX ""
foreach V of global X{
egen `V'_std=std(`V')
drop `V'
rename `V'_std `V'
global XX $XX `V'
}

* Recode the binary target variable 
recode foreign (0=1) (1=2)

* Save the initial dataset
save mydata , replace

* Split the initial dataset into a "training" and a "test" dataset
splitsample, generate(svar, replace) split(0.15 0.85) rseed(1) show

* Save the the training dataset as "data_train.dta" 
preserve
keep if svar==2
drop svar
save data_train , replace
restore

* Save the the test dataset as "data_test.dta" 
preserve
keep if svar==1
drop foreign svar
save data_test , replace
restore
********************************************************************************
\end{verbatim}
\end{stverbatim}
The outputs of this first part of the code are two datasets obtained through a random split of the initial \texttt{auto} dataset, \texttt{data\_train} used for in--sample fit,  and \texttt{data\_test} used for out--of--sample classification. It is important to bear in mind that the \texttt{data\_test} must contain only the (standardized) features, while both need to be without labels and missing values.   

In the second part of the code, we fit the learners to the training dataset. We perform this task by looping over the global \texttt{LEARNERS} that contains the name of the classifiers allowed by \texttt{c\_ml\_stata}.  

\begin{figure}[ht]
\centering
\includegraphics[width=14cm]{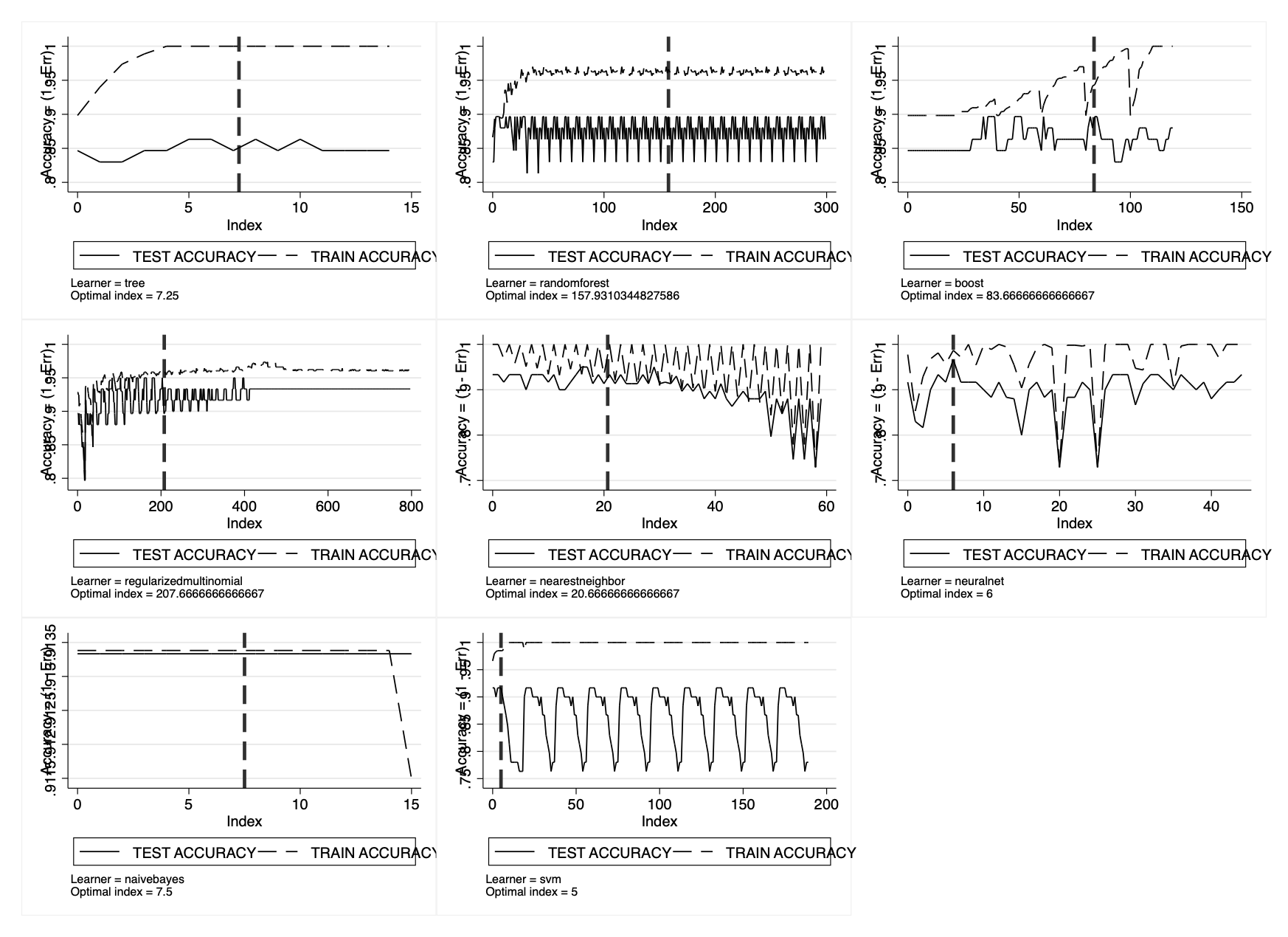}
\caption{Cross--validation maximum of the classification test accuracy over a greed of learners' tuning parameters. Accuracy measure: ``error rate'' (taking values between zero and one).}
\label{fig:fig4}
\end{figure}

\begin{stverbatim}
\begin{verbatim}
********************************************************************************
* FITTING ML CLASSIFIERS
******************************************************************************** 
* Define global macros Y, X, and LEARNERS for the fitting
global Y "foreign"
global X $XX
global LEARNERS tree randomforest boost ///
regularizedmultinomial nearestneighbor     ///
neuralnet naivebayes svm"

* Define a matrix M that will contain the main results
local i=1
global m: word count $LEARNERS
mat M=J($m,3,.)

* Fit all the ML methods by a loop over "LEARNERS" and save the main fitting
* results into the matrix M
foreach L of global LEARNERS{
* Load the training dataset
use data_train , clear
* Fit the single learner
c_ml_stata $Y $X , mlmodel(`L') in_prediction("in_pred_`i'") cross_validation("CV_`i'") ///
out_sample("data_test") out_prediction("out_pred_`i'") seed(10) save_graph_cv("graph_cv_`i'")
* Save results into M
mat M[`i',1]=e(TRAIN_ACCURACY)
mat M[`i',2]=e(TEST_ACCURACY)
mat M[`i',3]=e(BEST_INDEX) 
mat colnames M = TRAIN_ACCURACY TEST_ACCURACY index
mat rownames M = `L'
local i=`i'+1
}

* Turn the information contained in M into a dataset called "RES.dta"
clear
svmat M , n(col)
gen Learner=""
local i=1
foreach L of global LEARNERS{
replace Learner="`L'" in `i'
local i=`i'+1
}
replace index=int(index)
save RES , replace

* Put all the cross-validation graphs into a global macro called "TOT" 
global TOT ""
forvalues i=1/8{
global TOT $TOT graph_cv_`i'.gph
}

* Combine the graphs export the final graph
graph combine $TOT , scale(*0.5) plotregion(style(none)) scheme(s1mono) 
graph export cv_graph_all.png , as(png) replace
********************************************************************************
\end{verbatim}
\end{stverbatim}
The main outputs of this part are in the dataset \texttt{RES} containing, for every learner, the optimal greed index, the training and test accuracy, and the graph \texttt{cv\_graph\_all } displaying the cross--validation maximum of the classification test accuracy (i.e., the minimum of the classification error) over a greed of learners' tuning parameters. This graph is visible in figure \ref{fig:fig4}.

\begin{figure}[ht]
\centering
\includegraphics[width=14cm]{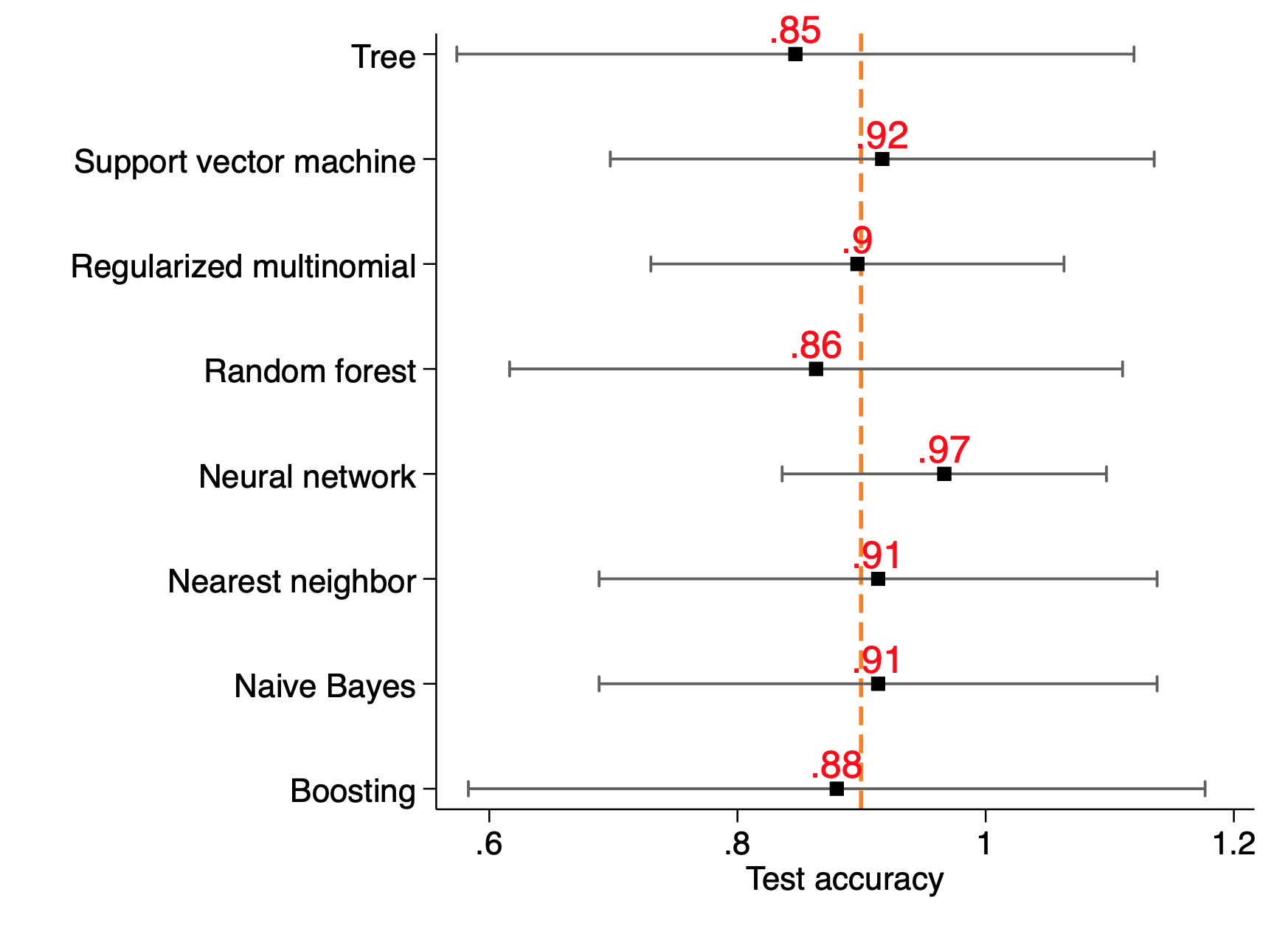}
\caption{Forest plot for comparing mean and standard deviation of different learners. Classification setting.}
\label{fig:fig5}
\end{figure}

\begin{stverbatim}
\begin{verbatim}
********************************************************************************
* FOREST PLOT FOR LEARNERS' ACCURACY
********************************************************************************
* Build a Forest Plot of the accuracy by Learner: mean & standard deviation 
clear
global LEARNERS tree randomforest boost ///
regularizedmultinomial nearestneighbor     ///
neuralnet naivebayes svm
global CV ""
forvalues i=1/8{
	use CV_`i'
	local L: word `i' of $LEARNERS
	cap drop Learner
	gen Learner="`L'"
	save CV_`i' , replace
	global CV $CV CV_`i'
}
********************************************************************************
clear
append using $CV
tab Learner , mis
********************************************************************************
merge 1:1 index Learner using RES
keep if _merge==3
sort Learner
********************************************************************************
sum mean_test_score 
global MEAN=r(mean)
********************************************************************************
replace Learner="Boosting" in 1
replace Learner="Naive Bayes" in 2
replace Learner="Nearest neighbor" in 3
replace Learner="Neural network" in 4
replace Learner="Random forest" in 5
replace Learner="Regularized multinomial" in 6
replace Learner="Support vector machine" in 7
replace Learner="Tree" in 8
********************************************************************************
cap drop C
gen C=`"""'
cap drop A
gen A=_n
tostring A,replace
cap drop _L
gen _L= A + " " + C + Learner + C
********************************************************************************
levelsof _L , local(XX) clean
global XX `XX'
********************************************************************************
cap drop _id
gen _id=_n
cap drop lo
gen lo=mean_test_score-1.96*std_test_score
cap drop hi
gen hi=mean_test_score+1.96*std_test_score
********************************************************************************
format mean_test_score %12.2g
********************************************************************************
twoway (rcap lo hi _id , horizontal ) (scatter  _id mean_test_score , ///
msymbol(S) msize(small) mcolor(black) mlabel(mean_test_score) ///
mlabposition(12) mlabc(red) mlabs(medlarge)) , ylabel($XX , angle(0)) ytitle("") ///
plotregion(style(none)) scheme(s1mono) xtitle("") legend(off) ///
xline($MEAN , lc(orange) lpattern(dash) lw(medthick) ) xtitle(Test accuracy)
graph export forestplot.png , as(png) replace
********************************************************************************
\end{verbatim}
\end{stverbatim}
The first part of this code extracts from the cross-validation results (contained, for every learner, in the dataset \texttt{CV\_}) the standard error of the test accuracy. The resulting dataset is then merged with the \texttt{RES} dataset, thus allowing for plotting mean and standard deviation of each learner's test accuracy within a forest plot. This graph is visible in figure \ref{fig:fig5}.
The results clearly show that in this dataset the neural network performs particularly well, with a cross--validation test accuracy of $0.97$ and the smallest confidence interval. One contrary, classification tree perform poorly, with a test accuracy of $0.85$ and a large confidence interval.  The other classifiers perform within this range and generally set out rather large confidence intervals. 

\section{Application 3: ML regression} \label{sec:app3}

In this application, we present the results from running a regression predictive analysis using \texttt{r\_ml\_stata}. For the sake of brevity, we do not report the code as it is identical to the one of the classification exercise, except for choosing as target the continuous variable \texttt{price} instead of the binary \texttt{foreign} (now included in the model as feature). 

\begin{figure}[ht]
\centering
\includegraphics[width=14cm]{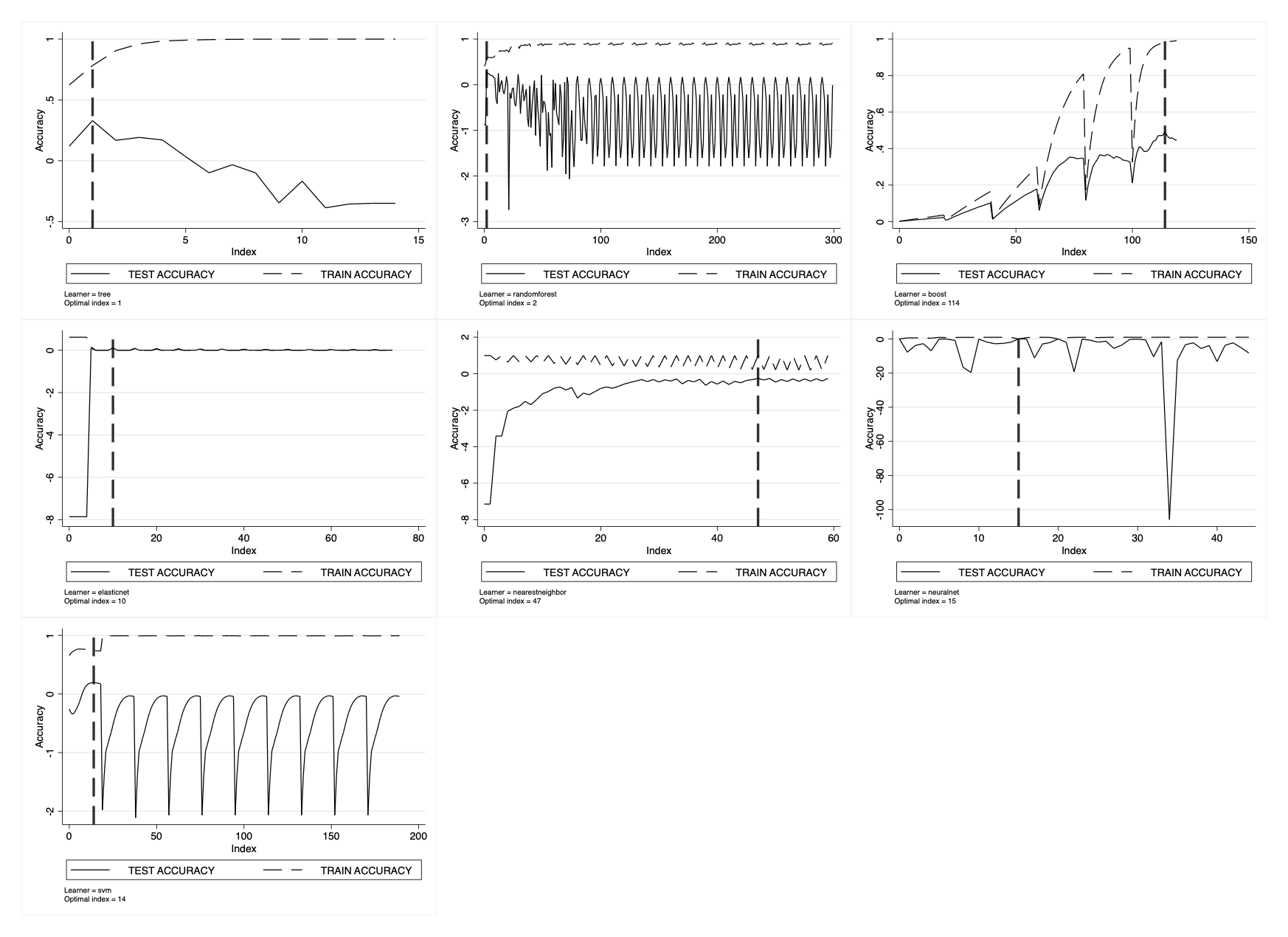}
\caption{Cross--validation maximum of the regression test accuracy over a greed of learners' tuning parameters. Accuracy measure: ``explained variance'' (taking one as maximum value). }
\label{fig:fig6}
\end{figure}

The cross--validation results are visible in figure \ref{fig:fig6} showing the greed index maximizing the test prediction accuracy. For regression, the selected accuracy measure is the prediction ``explained variance'' that takes a maximum value of one in the case of perfect prediction. 

\begin{figure}[ht]
\centering
\includegraphics[width=14cm]{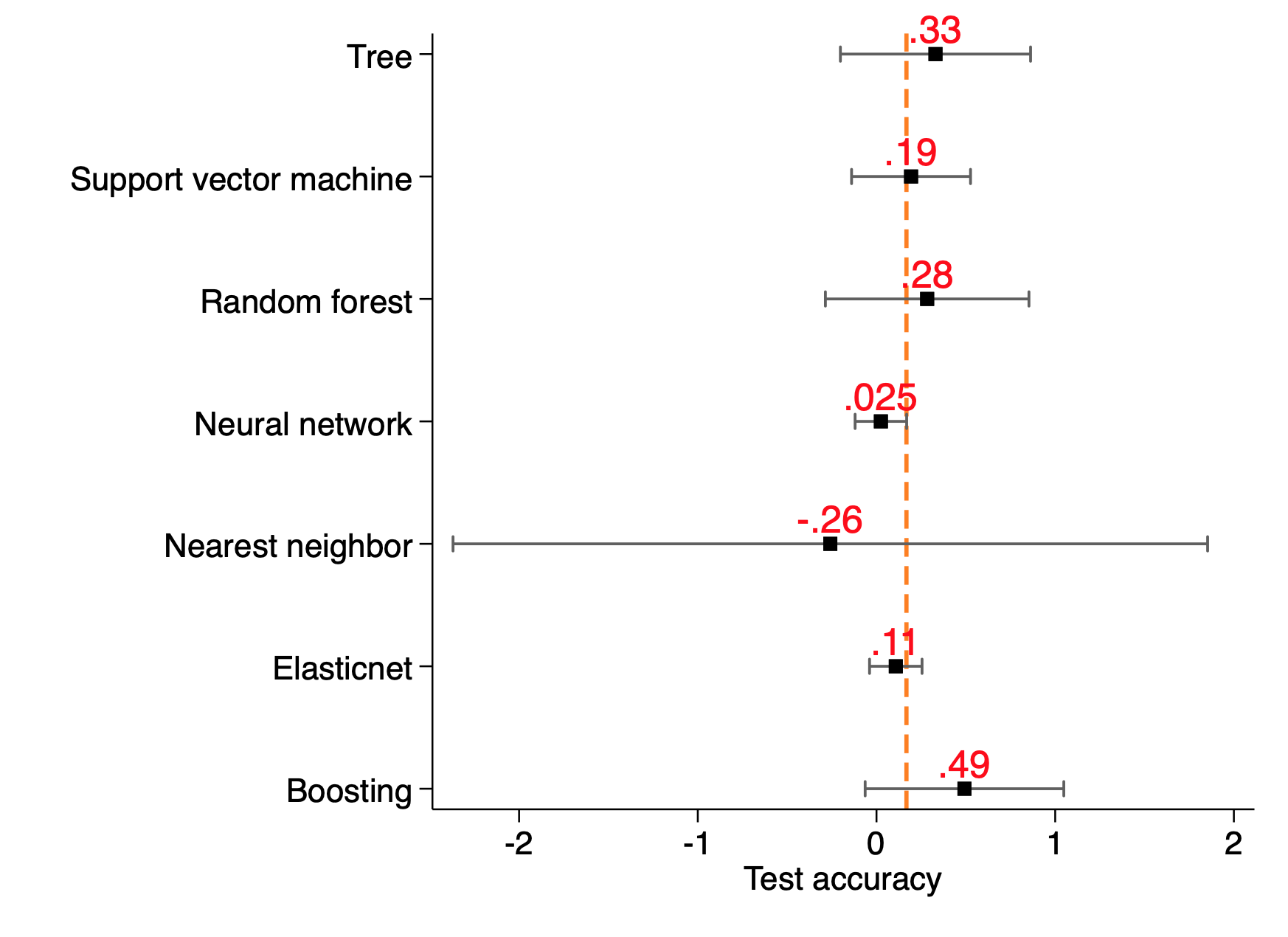}
\caption{Forest plot for comparing mean and standard deviation of different learners. Regression setting.}
\label{fig:fig7}
\end{figure}

Figure \ref{fig:fig7} sets out the comparison among the different learners computed by \texttt{r\_ml\_stata} in terms of the mean accuracy and standard deviation. Similarly to the classification case, we can observe that the learners behave differently, with boosting showing the best mean test accuracy ($0.49$) and tighter confidence interval. The worst performing learner is the nearest neighbor, both in term of mean accuracy and standard deviation.   

\section{Conclusion} \label{sec:concl}
In the last two decades, advances in statistical learning and computation have radically improved the prediction performance of targeted outcomes in pretty all scientific domains, including engineering, robotics, and artificial intelligence. ML has emerged as a new scientific paradigm to model outcomes and design pragmatic architectures for reasoned decision--making in uncertain environments.  Thanks to the recent Stata/Python integration platform introduced within Stata 16, producing Stata routines able to fit ML regression and classification has become a relatively straightforward task. 
By exploiting this opportunity, this paper has presented two related Stata modules, \texttt{r\_ml\_stata} and \texttt{c\_ml\_stata}, for fitting  popular ML models both in a regression and a classification setting. These commands provide hyper-parameters' optimal tuning via $K$-fold cross-validation using greed search, by wrapping the Python Scikit-learn API to perform cross-validation and outcome/label prediction.

Compared to other popular statistical software, Stata has the advantage to be highly user--friendly and powerful for complex data management. Unfortunately, Stata has not yet embedded a built-in Machine Learning package, except for the Lasso (including also the Elastic-net). 

The two commands herein presented thus go in the direction to partly fill this gap by providing the Stata users with two simple but powerful commands for fitting various ML methods. Further development of this work may include to provide also deep--learning Stata routines by wrapping into Stata the Python platforms Keras and Tensorflow.
 
\newpage

\bibliographystyle{sj}
\nocite*{}
\bibliographystyle{plain}



\clearpage
\end{document}